\documentclass[12pt]{article}
\input epsf
\usepackage{fullpage}
\usepackage{graphicx}
\newcommand{\be}{\begin{equation}}
\newcommand{\ee}{\end{equation}}
\newcommand{\bphi}{\mbox{\boldmath $\phi$}}

\font\mybb=msbm10 at 11pt

\def\bb#1{\hbox{\mybb#1}}

\newcommand{\news}{\setcounter{equation}{0}}
\def\ben{\begin{equation}}
\def\een{\end{equation}}
\def\bea{\begin{eqnarray}}
\def\eea{\end{eqnarray}}
\begin{document}
\title{
%\begin{flushright}\ \vskip -2cm {\tiny{\em DRAFT}}\end{flushright}
\vskip 2cm Knots in the Skyrme-Faddeev model}
\author{Paul Sutcliffe\\[10pt]
{\em \normalsize Department of Mathematical Sciences,
Durham University, Durham DH1 3LE, U.K.}\\
{\normalsize Email: p.m.sutcliffe@durham.ac.uk}}
\date{May 2007}
\maketitle
\begin{abstract}
The Skyrme-Faddeev model is a modified sigma model in
three-dimensional space, which has string-like topological solitons
classified by the integer-valued Hopf charge. Numerical simulations
are performed to compute soliton solutions for Hopf charges up to
sixteen, with initial conditions provided by families of rational
maps from the three-sphere into the complex projective line. A large
number of new solutions are presented, including a variety of torus
knots for a range of Hopf charges. Often these knots are only local
energy minima, with the global minimum being a linked solution, but
for some values of the Hopf charge they are good candidates
for the global minimum energy solution. The computed energies are in
agreement with Ward's conjectured energy bound.
\end{abstract}

\newpage
\section{Introduction}\news

\ \quad Over thirty years ago Faddeev suggested \cite{Fa2} that in
three-dimensional space the $O(3)$ sigma model, modified by the
addition of a Skyrme term, should have interesting string-like
topological solitons stabilized by the integer-valued Hopf charge.
Ten years ago substantial interest was generated by the first
attempts at a numerical construction of such solitons \cite{FN, GH}
and the suggestion that minimal energy solitons might take the form
of knots \cite{FN}. This conjecture has been confirmed by numerical
results, which demonstrate that the minimal energy soliton with Hopf
charge seven is a trefoil knot \cite{BS5}. Substantial numerical
investigations by several authors \cite{FN,GH,BS5,HS,HS2,Wa5} has
produced a comprehensive analysis of solitons with Hopf charges from
one to seven, and it appears that the global energy minima have now
been identified for these charges, together with several other
stable soliton solutions that correspond to local energy minima.

For Hopf charges five and six the minimal energy solitons form
links, but so far the charge seven solution remains the only known
knot, even including local energy minima. It is therefore currently
unknown whether this charge seven trefoil knot is the only knotted
solution or if there are many other knots, perhaps of various types,
which arise for higher Hopf charges. It is this question which is
addressed in the present paper. Solitons with Hopf charges up to
sixteen are constructed numerically and it is found that a variety
of torus knots exist at various Hopf charges. Often these knots are
only local energy minima, with the global minimum being a link, but
for some values of the Hopf charge it appears that a knot is a good
candidate for the global minimum energy solution.

One of the difficulties in extending previous numerical studies to
higher Hopf charges is that, even at low charges, there are a number
of local energy minima with large capture basins. This makes it
difficult to fully explore the landscape of local energy minima and
hence determine the global minimum, with the severity of the
problem generally increasing as the charge increases. To overcome
this problem an analytic ansatz is employed which uses rational maps
from the three-sphere into the complex projective line. Although
this ansatz does not provide any exact solutions, it does allow the
construction of reasonable initial conditions for any torus knot for
a large range of charges, plus a wide selection of links and
unknots. It is therefore possible to start at a variety of locations
in field configuration space, in particular where one suspects that
a local energy minimum might be close by, and explore the energy
landscape around this point. Not only is this a fruitful approach
for finding new solutions, but it can also be used to provide strong
evidence that a solution of a particular type does not exist, by
starting with an initial condition approximating this configuration
and demonstrating that it changes dramatically under energy
relaxation.

An analysis is made of the new knotted and linked solutions and some
understanding is obtained regarding the charges at which particular
knots are likely to exist.

\section{The Skyrme-Faddeev model and low charge solitons}\news
\ \quad The Skyrme-Faddeev model involves a map $\bphi:
\bb{R}^3\mapsto S^2,$ which is realized as a real three-component
vector $\bphi=(\phi_1,\phi_2,\phi_3),$ of unit length,
$\bphi\cdot\bphi=1.$ As this paper is concerned only with static
solutions then the model can be defined by its energy \be
E=\frac{1}{32\pi^2\sqrt{2}}\int \partial_i\bphi\cdot\partial_i\bphi
+\frac{1}{2}(\partial_i\bphi\times\partial_j\bphi)
\cdot(\partial_i\bphi\times\partial_j\bphi)\ d^3x, \label{energy}
\ee where the normalization is chosen for later convenience. The
first term in the energy is that of the usual $O(3)$ sigma model and
the second is a Skyrme term, required to provide a balance under
scaling and hence allow solitons with a finite non-zero size.

Finite energy boundary conditions require that the field tends to a
constant value at spatial infinity, which is chosen to be
$\bphi(\infty)=(0,0,1)={\bf e}_3.$  This boundary condition
compactifies space to $S^3,$ so that the field becomes a map $\bphi:
S^3\mapsto S^2.$ Such maps are classified by $\pi_3(S^2)=\bb{Z},$ so
there is an integer-valued topological charge $Q,$ the Hopf charge,
which gives the soliton number. Unlike most theories with
topological solitons, the topological charge is not a winding number
or degree of a mapping. Rather, it has a geometrical interpretation
as a linking number of field lines, as follows. Generically, the
preimage of a point on the target two-sphere is a closed loop (or a
collection of closed loops) in the three-sphere domain obtained from
the compactification of $\bb{R}^3.$ Two loops obtained as the
preimages of any two distinct points on the target two-sphere are
linked exactly $Q$ times, where $Q$ is the Hopf charge.

The Hopf charge can be written as the integral over the three-sphere
domain of a charge density, but this density is non-local in the
field $\bphi,$ as the following construction demonstrates. Let
$\omega$ denote the area two-form on the target two-sphere and let
$F=\bphi^*\omega$ be its pull-back under $\bphi$ to the domain
three-sphere. The triviality of the second cohomology group of the
three-sphere implies that $F$ is an exact two-form, say $F=dA.$ The
Hopf charge is then given by integrating the Chern-Simons three-form
over the three-sphere as \be Q=\frac{1}{4\pi^2}\int_{S^3}F\wedge
A.\ee

The energy bound \be E\ge c\, Q^{3/4},\quad \mbox{where} \quad
c=\left(\frac{3}{16}\right)^{3/8}\approx 0.534\ee has been proved
\cite{VK,KR}, and it is known that the fractional power is optimal
\cite{LY}, though it is expected that the above value of the
constant $c$ is not. Motivated by a study of the Skyrme-Faddeev
model on a three-sphere with a finite radius, together with an
analogy with the Skyrme model, Ward has conjectured \cite{Wa4} that
the above energy bound holds with the value $c=1$ (hence the choice
of normalization factor for the energy in (\ref{energy})), but this
has not been proven. As discussed shortly, the current energy values
known for low charge solitons are in good agreement with Ward's
conjectured energy bound, and later it will be demonstrated that the
energies for higher charge solitons are too.

The simplest examples of Hopf solitons are axially symmetric and
their qualitative features may be described as follows. First,
consider the model (\ref{energy}) defined in two-dimensional space,
so that the boundary conditions now result in a compactification of
space from $\bb{R}^2$ to $S^2.$ The field is then a map
$\bphi:S^2\mapsto S^2$ and is classified by an integer $m\in\bb{Z}
=\pi_2(S^2),$ which is the usual winding number familiar from the
$O(3)$ sigma model itself. Strictly speaking, in order to have
two-dimensional solitons with a finite size in this model then a
potential term must also be included, upon which the two-dimensional
solitons are known as baby Skyrmions \cite{PSZ}, but this aspect is
not crucial for the discussion here. Returning to the
three-dimensional model, then a toroidal field configuration can be
formed by embedding the two-dimensional soliton in the normal slice
to a circle in space. The two-dimensional soliton has an internal
phase and this can be rotated through an angle $2\pi n,$ as it
travels around the circle once, where $n\in \bb{Z}$ counts the
number of twists. A field configuration of this type has Hopf charge
$Q=nm,$ and it will be denoted by ${\cal A}_{n,m},$ so the first
subscript labels the number of twists and the second is the winding
number of the two-dimensional soliton forming the loop.

The minimal energy $Q=1$ soliton is of the type ${\cal A}_{1,1}$ and
has been studied numerically by a number of authors 
\cite{GH,FN,BS5,HS,Wa5} using different algorithms. 
Taken together, these results suggest that the charge one soliton has
an energy of around $E=1.21,$ and this is part of a
general pattern where the minimal energy soliton exceeds Ward's
conjectured bound by around $20\%.$ The numerical computations
reported in this paper produce a charge one energy of $E=1.204,$ and
hence appear to underestimate the energy by an amount of the order
of $1\%.$ For low charge solitons this accuracy could be improved by
increasing the resolution of the numerical grid, but it is
computationally too expensive to increase the resolution for all the
large number of simulations reported here for large charges.
Moreover, although the energy is a slight underestimate for a given
soliton solution, the relative energies between two given solutions
are much more accurate than this (changing grid resolutions and
sizes suggests an accuracy of around $0.1\%$) and it is more
important to preserve the relative energies than the absolute energy
of any single solution.

The details of the numerical computations are similar to those of
Ref.\cite{BS5}, and full details of the related numerical code used
to study Skyrmions can be found in Ref.\cite{BS3}. Briefly, a finite
difference scheme is employed with a lattice spacing $\Delta x=0.1$
on a grid containing  $(151)^3$ lattice points, with the field fixed
to the vacuum value $\bphi={\bf e}_3$ on the boundary of the grid;
this is preferable to other methods of dealing with the finite
volume simulation domain, such as the one used in Ref.\cite{BS5}
which led to a more substantial underestimate of energies, though
again relative energies had a high accuracy. The energy minimization
algorithm proceeds by evolving second order in time dynamics and
periodically removing kinetic energy from the system whenever the
potential energy of the system begins to increase. It is based on
the approach employed in Ref.\cite{BS5}, but is computationally more
efficient since the second order dynamics is derived from  the
kinetic term of the sigma model only, rather than the full kinetic
term determined from the Lorentz invariant Lagrangian associated
with the energy (\ref{energy}). This subtle difference leads to a
more efficient algorithm since the leading term in the evolution
equations is now diagonal, so a costly matrix inversion can be
avoided.

A sensible definition of the position of a soliton is to identify
where the field is as far as possible from the vacuum field. A Hopf
soliton is therefore string-like, since the soliton's position is
defined to be the closed loop (or collection of loops) corresponding
to the preimage of the point $\bphi=-{\bf e}_3=(0,0,-1),$ which is
antipodal to the vacuum value on the target two-sphere. To visualize
a field configuration this position curve is plotted, though for
clarity a tube around this position is displayed, given by an
isosurface of the form $\phi_3=-1+\nu,$ where generally the value
$\nu=0.2$ is chosen. If a pictorial representation of the Hopf
charge is also required then a second preimage curve can also be
plotted (again as a tube), and the linking number of these two
curves inspected. There is no natural choice for this second
preimage value (${\bf e}_3$ is not a very useful choice) and in this
paper when a second preimage curve is displayed it corresponds to
the point $\bphi=(\sqrt{2\mu-\mu^2},0,-1+\mu)$ where $\mu=0.1.$ In what
follows this curve will be referred to as the linking curve, and
since $\mu$ is relatively small then the linking curve remains
reasonably close to the position curve, making it fairly easy to
inspect the linking number.

\begin{figure}
\begin{center}
%\leavevmode 
%\epsfxsize=16cm\epsffile{1to7.ps} 
\includegraphics[width=16cm]{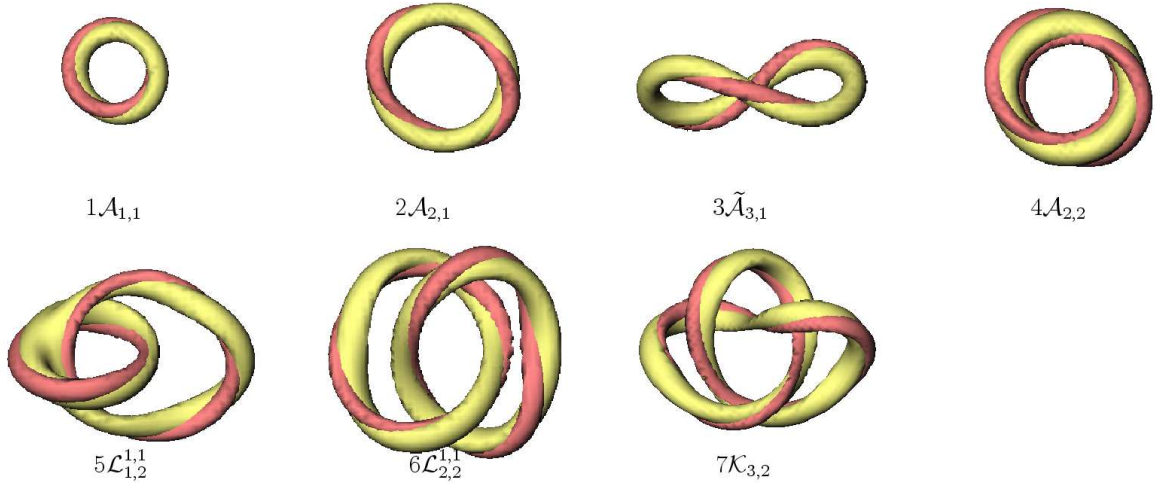} 
\caption{The position (light tube) and linking (dark tube)
curves for the known lowest energy solitons with Hopf
charges $1\le Q\le 7.$}
\label{fig-low}
\end{center}
\end{figure}

In Figure \ref{fig-low} the position (light tube) and linking (dark tube)
curves are displayed for the known lowest energy solitons with Hopf
charges from one to seven. Given the large number of numerical
simulations performed by different groups it seems reasonably
certain that these are the minimal energy solitons for these
charges. The energies of these solutions will be discussed later,
and as mentioned above they are generally around $20\%$ above Ward's
conjectured bound, but for now it is the structure of the solutions
that is of primary interest, and this is reviewed below.

The solutions with charges one and two are both axially symmetric,
and are of the type ${\cal A}_{1,1}$ and ${\cal A}_{2,1}$
respectively \cite{FN,GH}, with one and two twists of the linking
curve around the position curve clearly visible. There is a solution
of the form ${\cal A}_{1,2}$ \cite{Wa4,Wa5,HS} but it has an energy
around $13\%$ above that of the minimal charge two soliton. This
solution has the interpretation of two $Q=1$ solitons stacked one
above the other, preserving the axial symmetry.

 The charge three soliton is basically of the type ${\cal
A}_{3,1}$ but the axial symmetry is broken as the position curve is
bent \cite{BS5}, so it will be denoted by $\widetilde {\cal
A}_{3,1}$ to emphasize that the structure is deformed. There is an
axial solution of the type ${\cal A}_{n,1}$ for any integer $n,$ but
for $n>2$ this solution is unstable to a coiling instability
\cite{BS5}; a well-known phenomenon in other settings such as
twisted elastic rods. The $Q=4$ soliton is of the type ${\cal
A}_{2,2},$ \cite{HS2,Wa4}, and may be thought of as two $Q=2$ solitons
stacked one above the other. For both $Q=4$ and $Q=5$ there are also
bent solutions \cite{BS5}, similar to the charge three soliton, and
therefore denoted $\widetilde {\cal A}_{4,1}$ and $\widetilde {\cal
A}_{5,1},$ but unlike the $Q=3$ case, both these solutions are only 
local energy minima, with energies around
 $2\%$ and $4\%$ above the global minimum, respectively.
The minimal energy charge five soliton provides the first example of
a link, that is, a solution in which the position curve contains two
or more disconnected components. It consists of a charge one soliton
and a charge two soliton which are linked once \cite{HS2}, with the
result that each of these solitons gains an additional unit of
linking number to make a total of five. To denote a link of this
type the notation ${\cal L}_{1,2}^{1,1}$ will be used, where the
subscripts label the charges of the components of the links and the
superscript above each subscript counts the extra linking number of
that component, due to its linking with the others. Hence the total
charge is the sum of the subscripts plus superscripts. The charge
six soliton is similar to that of charge five,  but now both
components of the link have charge two \cite{BS5}, so using the
above notation it is written as ${\cal L}_{2,2}^{1,1}.$

Finally, the charge seven soliton is the first (and so far only)
example of a knot \cite{BS5}. The position curve is a trefoil knot,
which has a self-linking or crossing number of three (the crossing
number of a knot is not an invariant and its use in this paper
refers to the minimal crossing number over all presentations). An
examination of the linking curve in Figure \ref{fig-low} confirms
 that it twists
around the position curve four times as the knot is traversed, so
the Hopf charge is indeed seven, being the sum of the crossing
number plus the number of twists. Any field configuration of a
trefoil knot will be denoted by ${\cal K}_{3,2},$ which refers to
the fact that the trefoil knot is also the $(3,2)$-torus knot (see
the following section for more information regarding torus knots).
Note that this notation does not display the Hopf charge of the
configuration, which is determined by the twist number as well as
the crossing number, but this should not cause any confusion in what
follows. Of course, for knots and links, as well as for unknot
configurations, the precise conformation of the position curve is
important in determining the energy, not just its topological type,
but its knotted or linked structure is an important feature in
classifying the solution. Thus, for example, the $Q=7$ trefoil knot
displayed in Figure \ref{fig-low} is not as symmetric as its typical knot
theory presentation, and it is significant that the energy is
lowered by breaking the possible cyclic $C_3$ symmetry.

Having reviewed the known results for low charge solitons and
introduced the notation used to label their structural type, it is
time to turn to the main questions addressed in this paper. Given
the variety of solutions which appear, even at low charges, it is
difficult to make a confident prediction of the kind of behaviour
that might arise at higher charges. In particular, an interesting
open question is whether more knot solitons appear at higher charge,
and if so what types of knots arise and are they local or global
minima? It could have been the case that the charge seven trefoil
knot was the only knot soliton, though in this paper it will be
shown that this is far from true. Also, it would be useful to have
at least a qualitative understanding of the fact that $Q=7$ is the
lowest charge at which a knot first appears, and desirable to be
able to estimate the charges at which other knots might exist.

In this section it has been mentioned briefly that at each charge
there are often local minima in addition to the global minimum, and
it is expected that the number of local minima generally increases
with the charge. As noted in the introduction, these local minima
can have large capture basins, making it difficult to fully explore
the space of solutions. For any relaxation computation this makes
the choice of initial conditions a crucial issue. In particular, it
is not obvious how to construct a variety of 
reasonably low-energy initial field
configurations for all Hopf charges. One family of configurations
are the axial fields ${\cal A}_{n,1},$ which are all unstable for
$n>2.$ However, even for charges as low as four, initial conditions
which use small perturbations of these unstable solutions get
trapped in local minima. One successful approach for low charges
involves constructing links by hand \cite{HS,HS2}, using a numerical
cut-and-paste technique where various numerical field configurations
are sewn together in different parts of the simulation grid to
create a linked field with a given Hopf charge. Although this
approach was the first method to successfully yield the global
minima at charges four and five, it becomes more cumbersome for
larger charges and is not applicable for creating knot initial
conditions; it is also not very elegant mathematically, but this is
perhaps not a serious criticism. Another set of initial conditions
that have been used \cite{HS2} are based on the fields ${\cal
A}_{n,m},$ with $m>1,$ though these fields tend to have relatively
large energies and therefore require quite long simulation times
(except for the low charge examples of ${\cal A}_{1,2}$ and ${\cal
A}_{2,2}).$ In the following section an analytic ansatz is presented
that yields reasonable field configurations for a variety of charges
and describes all torus knots plus a wide range of links, in addition
to fields of the type ${\cal A}_{n,m}.$ The key ingredient of the ansatz
involves a rational map from the three-sphere into the complex
projective line. The ansatz allows the construction of initial
conditions corresponding to a variety of locations in field
configuration space; in particular, configurations can be created
which are reasonably close to suspected local energy minima. This
approach substantially reduces the computational effort required to
explore the energy landscape and makes it feasible to study knots
and links for quite large charges.

\section{Rational maps, torus knots and links}\news
\ \quad Torus knots are classified by a pair $(a,b)$ of coprime
positive integers with $a>b.$ A knot is a torus knot if it lies on
the surface of a torus, in which case the integers $a$ and $b$ count
the number of times that the knot winds around the two cycles of the
torus. The simplest example is the trefoil knot, which is the
$(3,2)$-torus knot and has three crossings. In the standard knot
catalogue notation it is $3_1,$ where the number refers to the
crossing number of the knot and the subscript labels its position in
the knot catalogue. Other torus knots which will be of interest in
this paper include the $(5,2)$-torus knot, also known as Solomon's
seal knot, which has five crossings (knot $5_1$ in the catalogue),
and the $(4,3)$-torus knot which has crossing number eight (knot
$8_{19}$ in the catalogue). In general the $(a,b)$-torus knot has
crossing number $C=a(b-1).$

Denote by ${\cal K}_{a,b}$ any field configuration in which the
position curve is an $(a,b)$-torus knot. Such a field configuration
will have Hopf charge $Q=C+T,$ where $C=a(b-1)$ is the knot crossing
number and the integer $T$ counts the number of times that the
linking curve twists around the position curve (only situations
where the orientation is such that $T$ is non-negative will be of
relevance here). The main aim of this section is to construct fields
of type ${\cal K}_{a,b}$ for a large range of charges $Q.$

The first step is to recall the standard description \cite{BK} of a
torus knot as the intersection of a complex algebraic curve with the
three-sphere. Consider $(Z_1,Z_0)\in \bb{C}^2$ and the unit
three-sphere $S^3\subset \bb{C}^2$ given by $|Z_1|^2+|Z_0|^2=1.$
Then the intersection of this three-sphere with the complex
algebraic curve $Z_1^a+Z_0^b=0$ is indeed the $(a,b)$-torus knot.

To use the above description to produce a field configuration, the
spatial coordinates $(x_1,x_2,x_3)\in \bb{R}^3$ are mapped to the
unit three-sphere via a degree one spherically equivariant map.
Explicitly, \be (Z_1,Z_0)=((x_1+ix_2)\frac{\sin f}{r},\cos
f+i\frac{\sin f}{r}x_3),\ee where $r^2=x_1^2+x_2^2+x_3^2$ and the
profile function $f(r)$ is a monotonically decreasing function of
the radius $r$, with boundary conditions $f(0)=\pi$ and
$f(\infty)=0.$ The precise form of this function will be specified
shortly.

A Riemann sphere coordinate $W,$ is used on the target two-sphere of
the field $\bphi,$ determined via stereographic projection as \be
W=\frac{\phi_1+i\phi_2}{1+\phi_3}.\ee With these definitions, then
constructing a map $\bphi:\bb{R}^3\mapsto S^2$ is equivalent to
specifying $W(Z_1,Z_0),$ that is, $W:S^3\mapsto \bb{C}\bb{P}^1,$ a
map from the three-sphere to the complex projective line. This map
will be taken to be a rational map, which means that $W=p/q,$ where
both $p$ and $q$ are polynomials in the variables $Z_1$ and $Z_0.$

For an $(a,b)$-torus knot the mapping is chosen to have the form \be
W=\frac{p}{q}=\frac{Z_1^\alpha Z_0^\beta}{Z_1^a+Z_0^b},
\label{ratmap}\ee where $\alpha$ is a positive integer and $\beta$
is a non-negative integer. A map of this form has the correct
boundary conditions at spatial infinity, since as
$r\rightarrow\infty$ then $(Z_1,Z_0)\rightarrow(0,1)$ and, because
$\alpha>0,$ this gives $W\rightarrow 0,$ which is the stereographic
projection of the point $\bphi=(0,0,1)={\bf e}_3.$ Furthermore, the
position curve is the preimage of the point $-{\bf e}_3,$ which
corresponds to $W=\infty,$ and hence is given by $q=0=Z_1^a+Z_0^b.$
As the coordinates $(Z_1,Z_0)$ automatically lie on the unit
three-sphere then the position curve is the $(a,b)$-torus knot, for
all values of $\alpha$ and $\beta.$

The Hopf charge of the map determined by (\ref{ratmap}) is given by
the cross ratio \be Q=\alpha b+\beta a.\label{charge}\ee To see this
note that a rational map \be W:S^3\mapsto \bb{C}\bb{P}^1, \quad
\mbox{given\ by} \quad W(Z_1,Z_0)=\frac{p(Z_1,Z_0)}{q(Z_1,Z_0)},\ee
has a natural extension to a map \be (p,q):\bb{B}^4\subset
\bb{C}^2\mapsto \bb{C}^2, \quad \mbox{given\ by} \quad
(p(Z_1,Z_0),q(Z_1,Z_0)).\ee Here $\bb{B}^4$ denotes the 4-ball with
boundary  $S^3,$ and is obtained by replacing the constraint
$|Z_1|^2+|Z_0|^2=1$ by the inequality $|Z_1|^2+|Z_0|^2\le 1.$ The
map $(p,q)$ is between manifolds of the same dimension and 
has a degree, defined by counting preimages of a generic point of
the target space, weighted by the signs of the Jacobian. As all the
other mappings involved in the ansatz have degree one, and the
standard Hopf map $(p,q)\mapsto p/q$ is employed, then the Hopf
charge $Q$ of the combined mapping is equal to the degree of the
mapping $(p,q).$ Furthermore, as the mapping $(p,q)$ is holomorphic
then the sign of the Jacobian of all generic preimage points is
positive, so the degree is simply the number of preimages.

For the mapping $(p,q)$ given by (\ref{ratmap}) take the generic
point in target space to be $(\epsilon,0),$ so the Hopf charge is
the number of solutions of the equation $(Z_1^\alpha
Z_0^\beta,Z_1^a+Z_0^b)=(\epsilon, 0).$ Writing $Z_0$ in terms of
$Z_1$ from the first component of this equation and substituting
into the second component yields the polynomial equation
$Z_1^{\alpha b+\beta a}=(-1)^\beta\epsilon^b,$ which clearly has
$Q=\alpha b+\beta a$ solutions, as stated.

For any $(a,b)$-torus knot the above ansatz can be used to construct
a field configuration of the type ${\cal K}_{a,b}$ with Hopf charge
$Q,$ providing there are integers $(\alpha,\beta),$ with $\alpha$
positive and $\beta$ non-negative, for which $Q=\alpha b+\beta a.$
For example, in the case of the trefoil knot then $(a,b)=(3,2)$ hence any
$Q>3$ can be obtained. In fact a trefoil knot with $Q=3$ can also be
obtained by interchanging the roles of $Z_1$ and $Z_0,$ but a field
of this type is not relevant, as will be made clear later. Note that
for some values of $Q,$ and fixed $(a,b),$ there are multiple
solutions for $(\alpha,\beta).$ These multiple solutions produce
qualitatively similar fields, in that both describe the same
$(a,b)$-torus knot position curve, and the linking curve twists the
same number of times around the position curve; though the
distribution of twist may vary. In examples where such multiple
solutions were used as initial conditions, the energy relaxation
algorithm produced identical final results for all choices.

Recall that a profile function $f(r)$ needs to be specified,
satisfying the boundary conditions $f(0)=\pi$ and $f(\infty)=0.$ In
fact, these are the correct boundary conditions in an infinite
domain but in the finite domain used for numerical simulations the
last boundary condition needs to be replaced by $f=0$ at the
boundary of the numerical grid. As the ansatz is only used to
provide initial conditions then most reasonable monotonic functions
will suffice. The simulations discussed in this paper were performed
on a cubic grid of side-length $2L,$ that is, $-L\le x_i\le L,$
where $2L=15.$ A simple linear profile function was used, given by
$f(r)=\pi(L-r)/L,$ for $r\le L$ and zero otherwise. For a given
rational map one could aim to minimize the energy of the ansatz over
all profile functions, but this has not been attempted for reasons
discussed shortly.

\begin{figure}
\begin{center}
%\leavevmode 
%\epsfxsize=10cm\epsffile{approx76.ps}
\includegraphics[width=10cm]{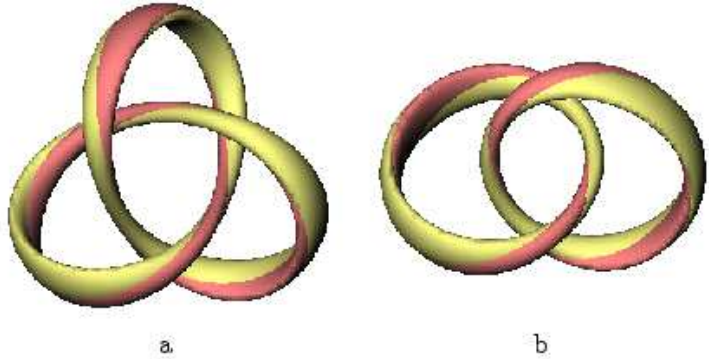}  
\caption{The position (light tube) and linking (dark tube)
curves for initial conditions created using the rational map
ansatz: (a) $Q=7$ trefoil knot; (b) $Q=6$ link 
${\cal L}_{2,2}^{1,1}.$ }
\label{fig-approx}
\end{center}
\end{figure}

As an illustration of the use of the ansatz, consider an
approximation to the minimal energy $Q=7$ trefoil knot. To construct
a field ${\cal K}_{3,2}$ with $Q=2\alpha+3\beta=7,$ requires
$(\alpha,\beta)=(2,1),$ resulting in the map
$W=Z_1^2Z_0/(Z_1^3+Z_0^2).$ The field generated from this rational
map is displayed in Figure \ref{fig-approx}a, 
where both the position and linking
curves are shown. It is clear from this figure that the field has
similar qualitative features to the minimal energy soliton, in that
it is a trefoil knot with four twists. The main qualitative
difference is that the ansatz produces a more symmetric field, having a
cyclic $C_3$ symmetry. This symmetry is obvious from the form of the
rational map, as a rotation by $120^\circ$ around the $x_3$-axis
corresponds to the transformation $Z_1\mapsto e^{2\pi i/3}Z_1,$ upon
which the rational map changes by only a phase; which is simply an
action of the global $O(2)$ symmetry on target space.

Using the above field as the initial condition for an energy
relaxation simulation yields the minimal energy soliton very
quickly, and certainly requires far less computational resources
than computing the solution from a perturbed axial configuration.
Note that, in practice, the cyclic $C_3$ symmetry is slightly broken
by the cubic numerical grid, and in particular its boundary, so an
explicit symmetry breaking perturbation is not required. In examples
discussed later, where initial conditions have a cyclic $C_4$
symmetry, then the configuration is created with a slight
displacement from the centre of the grid to again slightly break any
exact symmetry.

As mentioned above, the profile function could be optimized to find
the minimal energy configuration, for a given rational map, but this
has not been attempted. As with the above example, generally the
ansatz is more symmetric than the relaxed solution, so it is not
expected that a good estimate of the energy can be obtained in this
way. It might be possible to improve the approximation to a
quantitative level by minimizing over families of rational maps,
which include symmetry breaking terms, rather than using the
rational maps in this paper; chosen to be the simplest with the
correct qualitative features. In this case it would be worthwhile to
also minimize the profile function, but as the minimization problem
couples the rational map and the profile function together then this
is not a simple numerical task. In fact, this numerical problem
appears to have about the same level of difficulty as performing
energy relaxation via full field simulations. As the energy
relaxation code produces a solution very quickly and efficiently
from the simple ansatz, then there appears little need to try and
improve upon it. The profile function could be adjusted to a minimal
extent, for example, by including parameters to set the scale and
thickness of the knot, but again this has not been necessary.

So far only knot initial conditions have been considered, but the
ansatz can also be used to construct a variety of unknots and links
by choosing suitable rational maps. For example, axial fields of the
form ${\cal A}_{n,m}$ are generated by a rational map
$W=Z_1^n/Z_0^m.$

Links correspond to rational maps in which the curve determined by
the denominator is reducible. The torus knot maps with $(a,b)$
coprime degenerate to links if $(a,b)$ are not coprime. As an
example, links of the type ${\cal L}_{n,n}^{1,1}$ (which have Hopf
charge $Q=2n+2$) are derived from the rational map \be
W=\frac{Z_1^{n+1}}{Z_1^2-Z_0^2} =
\frac{Z_1^n}{2(Z_1-Z_0)}+\frac{Z_1^n}{2(Z_1+Z_0)},\label{links}\ee
where the partial fraction decomposition reveals the charges of the
constituent links by comparison with the axial maps. The $Q=6$
field constructed using the rational map $(\ref{links})$ with $n=2$
is displayed in Figure \ref{fig-approx}b.
 As can be seen from this figure, the
field is qualitatively similar to the minimal energy $Q=6$ soliton,
which is of the type ${\cal L}_{2,2}^{1,1}.$ More examples of
rational maps associated to linked configurations will appear in the
following section.

\begin{table}
%\centering
\vskip -2cm
\begin{tabular}{|c|c|c|c|c|c|c|}
\hline
 & \multicolumn{6}{|c|}{initial} \\
Q & \multicolumn{6}{|c|}{$\downarrow$} \\
 & \multicolumn{6}{|c|}{final} \\
\hline
 & ${\cal K}_{3,2}$& & & & &\\
5 & $\downarrow$& & & & &\\
  &  ${\cal L}_{1,2}^{1,1}$& & & & &\\ \hline
 & ${\cal K}_{3,2}$& & & & &\\
6 & $\downarrow$& & & & &\\
  &  ${\cal L}_{2,2}^{1,1}$& & & & &\\ \hline
 & ${\cal K}_{3,2}$&${\cal L}_{2,3}^{1,1}$ & & & &\\
7 & $\downarrow$&$\downarrow$ & & & &\\
  &  ${\cal K}_{3,2}$&${\cal K}_{3,2}$ & & & &\\ \hline
 &${\cal L}_{3,3}^{1,1}$  &
${\cal L}_{2,2}^{2,2}$  &${\cal L}_{4,2}^{1,1}$ & ${\cal K}_{5,2}$ &
${\cal K}_{3,2}$
& \\
8 & $\downarrow$&$\downarrow$ &$\downarrow$ &$\downarrow$
 &$\downarrow$ &\\
 &${\cal L}_{3,3}^{1,1}$
 &${\cal L}_{3,3}^{1,1}$ &${\cal L}_{3,3}^{1,1}$
&${\cal L}_{3,3}^{1,1}$ &${\cal K}_{3,2}$ & \\ \hline
& ${\cal L}_{1,1,1}^{2,2,2}$& ${\cal K}_{4,3}$ & ${\cal
K}_{3,2}$&${\cal K}_{5,2}$
&${\cal L}_{2,3}^{2,2}$  &${\cal L}_{3,4}^{1,1}$\\
9 & $\downarrow$&$\downarrow$ &$\downarrow$ &$\downarrow$
&$\downarrow$
 &$\downarrow$\\
 &${\cal L}_{1,1,1}^{2,2,2}$ &${\cal L}_{1,1,1}^{2,2,2}$
 &  ${\cal K}_{3,2}$&${\cal K}_{3,2}$ &${\cal K}_{3,2}$&${\cal K}_{3,2}$
\\ \hline
& ${\cal L}_{1,1,2}^{2,2,2}$& ${\cal K}_{4,3}$ & ${\cal
L}_{3,3}^{2,2}$ &${\cal L}_{5,3}^{1,1}$
&${\cal K}_{5,2}$  &${\cal K}_{3,2}$\\
10 & $\downarrow$&$\downarrow$ &$\downarrow$ &$\downarrow$
&$\downarrow$
 &$\downarrow$\\
 &${\cal L}_{1,1,2}^{2,2,2}$ &${\cal L}_{1,1,2}^{2,2,2}$
 &  ${\cal L}_{3,3}^{2,2}$& ${\cal L}_{3,3}^{2,2}$ & ${\cal L}_{3,3}^{2,2}$
 &${\cal K}_{3,2}$
\\ \hline
& ${\cal K}_{4,3}$ & ${\cal K}_{5,2}$ & ${\cal K}_{7,2}$ &${\cal
K}_{3,2}$
& &\\
11 & $\downarrow$&$\downarrow$ &$\downarrow$ &$\downarrow$ &
 &\\
 &${\cal L}_{1,2,2}^{2,2,2}$ & ${\cal K}_{5,2}$
 &  ${\cal L}_{3,4}^{2,2}$& ${\cal K}_{3,2}$& &
\\ \hline
& ${\cal K}_{5,3}$ & ${\cal K}_{3,2}$ & ${\cal K}_{5,2}$ &${\cal
K}_{7,2}$
& &\\
12 & $\downarrow$&$\downarrow$ &$\downarrow$ &$\downarrow$ &
 &\\
 &${\cal L}_{2,2,2}^{2,2,2}$ & ${\cal K}_{4,3}$
 &  ${\cal K}_{5,2}$& ${\cal L}_{4,4}^{2,2}$& &
\\ \hline
& ${\cal K}_{3,2}$ & ${\cal K}_{5,3}$ & ${\cal K}_{5,2}$ &${\cal
K}_{7,2}$
& &\\
13 & $\downarrow$&$\downarrow$ &$\downarrow$ &$\downarrow$ &
 &\\
 &${\cal K}_{4,3}$ & ${\cal X}_{13}$
 &  ${\cal K}_{5,2}$& ${\cal L}_{3,4}^{3,3}$& &
\\ \hline
& ${\cal K}_{4,3}$ & ${\cal K}_{5,3}$ &${\cal K}_{3,2}$  & & &\\
14 & $\downarrow$&$\downarrow$ &$\downarrow$ & &
 &\\
 &${\cal K}_{4,3}$ & ${\cal K}_{5,3}$ & ${\cal K}_{5,2}$  & & &
\\ \hline
& ${\cal K}_{5,3}$ & ${\cal K}_{4,3}$ & ${\cal K}_{3,2}$  & & &\\
15 & $\downarrow$&$\downarrow$ &$\downarrow$ & &
 &\\
 &${\cal X}_{15}$ & ${\cal L}_{1,1,1}^{4,4,4}$ & ${\cal K}_{5,3}$ & & &
\\ \hline
& ${\cal K}_{4,3}$ & ${\cal L}_{1,1,1,1}^{3,3,3,3}$ & ${\cal K}_{3,2}$ & & &\\
16 & $\downarrow$&$\downarrow$ &$\downarrow$ & &
 &\\
 &${\cal X}_{16}$ & ${\cal X}_{16}$  &${\cal X}_{16}$  & & &
\\ \hline
\end{tabular}
%\begin{flushleft}\caption{Caption}\end{flushleft}
\vskip 1cm \ \hskip -0cm Table 1: Initial conditions and final
solutions.
 \label{tab-relax}
\end{table}
\begin{table}
\vskip -26.7cm \ \hskip 10cm
%\centering
\begin{tabular}{|c|c|c|c|}
\hline $Q$ &  type & $E$ & $E/Q^{3/4}$\\  \hline 1 & ${\cal
A}_{1,1}$ &  1.204 &  1.204\\ \hline 2 & ${\cal A}_{2,1}$ &  1.967 &
1.170\\ \hline 3 & $\widetilde{\cal A}_{3,1}$ &  2.754 &  1.208\\
\hline 4 & ${\cal A}_{2,2}$ &  3.445 & 1.218 \\ \hline 5 & ${\cal
L}_{1,2}^{1,1}$ &  4.095 & 1.225 \\ \hline 6 & ${\cal
L}_{2,2}^{1,1}$ & 4.650  & 1.213 \\ \hline 7 & ${\cal K}_{3,2}$ &
5.242  & 1.218 \\ \hline
8 & ${\cal L}_{3,3}^{1,1}$ & 5.821  & 1.224 \\
8 & ${\cal K}_{3,2}$ &5.824  & 1.224 \\ \hline
9  & ${\cal L}_{1,1,1}^{2,2,2}$ &  6.360 & 1.224 \\
9  & ${\cal K}_{3,2}$ & 6.385 &  1.229 \\ \hline
10  & ${\cal L}_{1,1,2}^{2,2,2}$ &  6.905 & 1.228 \\
10  & ${\cal L}_{3,3}^{2,2}$ & 6.923  &  1.231 \\
10  & ${\cal K}_{3,2}$ &  6.973  &  1.240 \\ \hline
11  & ${\cal L}_{1,2,2}^{2,2,2}$ & 7.391 &  1.224 \\
11  & ${\cal K}_{5,2}$ &  7.502 & 1.242  \\
11  & ${\cal L}_{3,4}^{2,2}$ &  7.533 &  1.247 \\
11  & ${\cal K}_{3,2}$ &  7.614  &  1.261 \\ \hline
12  & ${\cal L}_{2,2,2}^{2,2,2}$ &7.833   &  1.215  \\
12  & ${\cal K}_{4,3}$ & 7.857 &  1.219   \\
12  & ${\cal K}_{5,2}$ & 8.070 &  1.252  \\
12  & ${\cal L}_{4,4}^{2,2}$ & 8.093 &   1.255 \\ \hline
13  & ${\cal K}_{4,3}$ & 8.272 &1.208  \\
13  & ${\cal X}_{13}$ & 8.462 &1.236  \\
13  & ${\cal K}_{5,2}$ &8.574  & 1.252 \\
13  & ${\cal L}_{3,4}^{3,3}$ & 8.633  &  1.261 \\  \hline
14  & ${\cal K}_{4,3}$ & 8.761  & 1.210  \\
14  & ${\cal K}_{5,3}$ &  8.807  &  1.217  \\
14  & ${\cal K}_{5,2}$ &  9.124  &  1.261 \\ \hline
15  & ${\cal X}_{15}$ & 9.290   & 1.219  \\
15  & ${\cal L}_{1,1,1}^{4,4,4}$ & 9.404  & 1.234  \\
15  & ${\cal K}_{5,3}$ & 9.408 & 1.234  \\ \hline 16  & ${\cal
X}_{16}$ & 9.769 &  1.221  \\ \hline
\end{tabular}
\vskip 1cm \hskip 9.5cm Table 2: Solution types and
energies.\label{tab-energies}
\end{table}

\section{Higher charge solitons}\news
\ \quad In this section the results of a large number of energy
minimization simulations are presented. Hopf charges up to sixteen
are studied, with initial conditions consisting of a variety of
links and knots, created using the rational map ansatz discussed
above.

\begin{figure}
\begin{center}
\leavevmode 
\includegraphics[width=12cm]{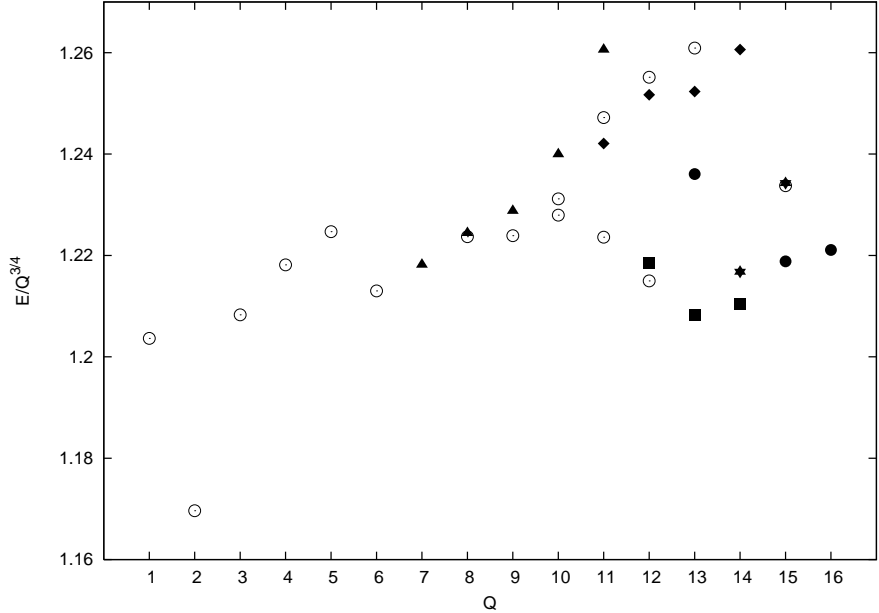}  
\caption{The ratio of the energy to the conjectured bound, that is, 
$E/Q^{3/4},$ as a function of  Hopf charge $Q$ for a variety of solutions:
unknots and links (white circles); ${\cal K}_{3,2}$ (triangles);
 ${\cal K}_{5,2}$ (diamonds); ${\cal K}_{4,3}$ (squares);
 ${\cal K}_{5,3}$ (stars); links which are not resolved (black circles).}
\label{fig-energy}
\end{center}
\end{figure}

Table 1 summarizes the simulations performed, by listing the type of
each initial condition together with the type of the resulting final
solution, obtained from the energy minimization algorithm. In Table
2 the types and energies of the known lowest energy solutions, plus
some local energy minima, are presented for $1\le Q\le 16,$ ordered
by increasing energies. Also given in Table 2 are the energies
divided by Ward's conjectured bound, $E/Q^{3/4},$ from which it can
be seen that the minimal energy solution is consistently around
$20\%$ above the conjectured bound. The values of $E/Q^{3/4}$ are
plotted in Figure \ref{fig-energy} using a notation that 
identifies the different
torus knots as follows. White circles denote configurations which do not
contain knots, that is, either unknots or links. The symbols
denoting knots are triangles for ${\cal K}_{3,2}$, diamonds for
${\cal K}_{5,2},$ squares for ${\cal K}_{4,3}$ and stars for ${\cal
K}_{5,3}.$ Configurations denoted by black circles will be discussed shortly.

\begin{figure}
\begin{center}
\leavevmode 
\includegraphics[width=14cm]{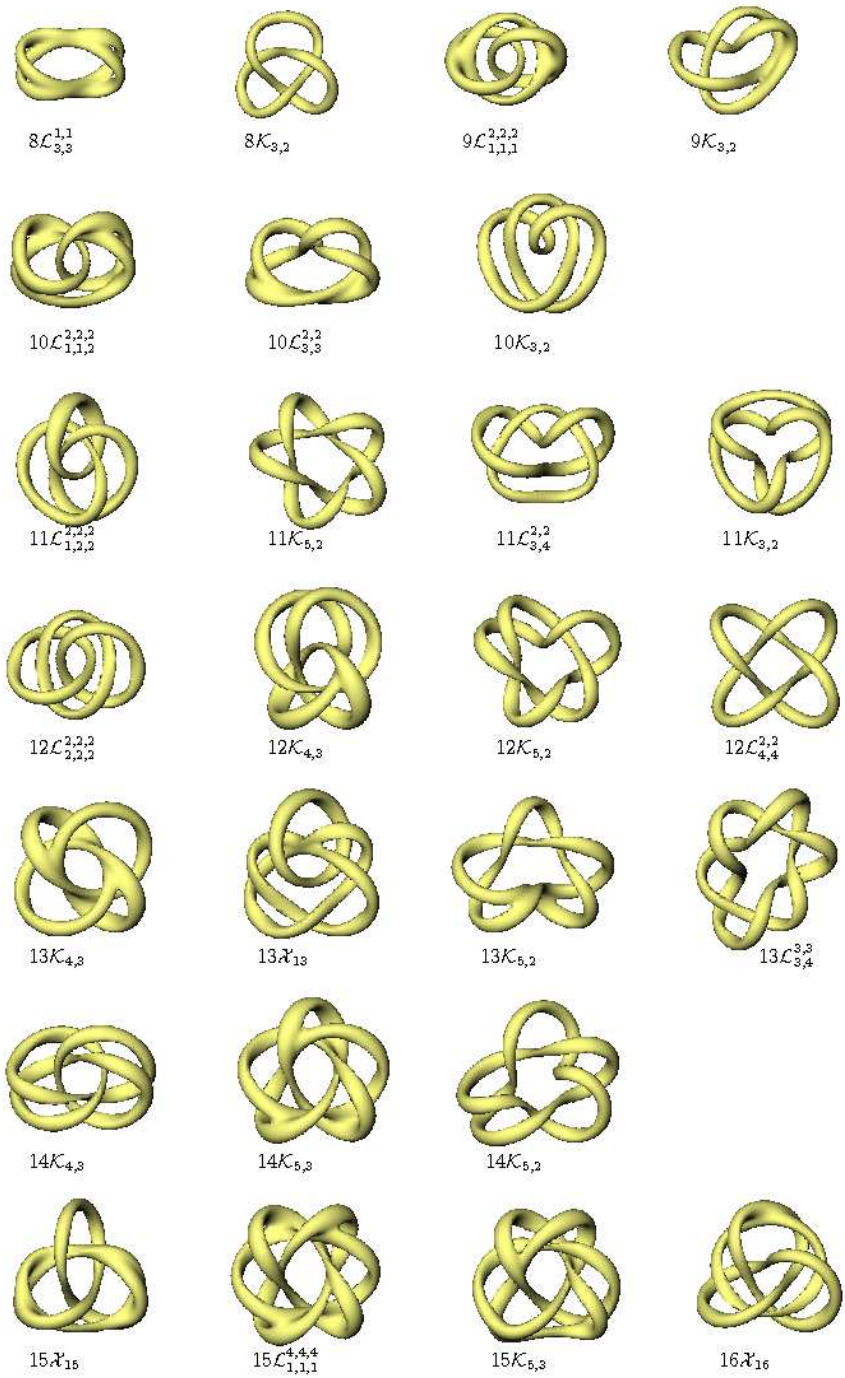} 
\caption{The position curves for a variety of solutions with Hopf charge
 $8\le Q\le 16.$ Each plot is labeled by its charge and type.}
\label{fig-high}
\end{center}
\end{figure}

 The present study has produced 26 new solutions with $8\le Q\le 16.$
The position curves for each of these is displayed in 
Figure \ref{fig-high}, but 
for clarity the linking curves are not shown,
although they have been examined to confirm the correct linking
number identifications. Each plot is labeled by its charge and type,
with energies increasing first from left to right and then top to
bottom.

First of all, consider trefoil knots, that is, solutions which have
the form ${\cal K}_{3,2}.$ As mentioned earlier, the $Q=7$ minimal
energy soliton has this form and is obtained from the related
rational map initial condition. This is encoded in Table 1 as the
process ${\cal K}_{3,2}\rightarrow{\cal K}_{3,2}.$ The entry to the
right of this one in Table 1 reveals that the same trefoil knot
solution is also obtained from the linked initial condition ${\cal
L}_{2,3}^{1,1}.$

There are no trefoil knots with $Q<7.$  Evidence supporting this is
presented in Table 1 for $Q=5$ and $Q=6,$ where it is seen that in
both cases initial conditions of the form ${\cal K}_{3,2}$ result in
the linked minimal energy solutions, which are ${\cal
L}_{1,2}^{1,1}$ and ${\cal L}_{2,2}^{1,1}$ respectively. A
reasonable interpretation of these results is that for $Q<7$ the
number of twists $T=Q-3$ is too low, given the preferred length of
the soliton in a trefoil knot arrangement. In other words, the twist
per unit length is too low to be an energetically efficient
distribution of the Hopf charge between crossing and twisting. A
more detailed discussion of this aspect will be given later, when
general torus knots will be considered. Note that, in particular,
there is no trefoil knot solution in which there is no twist, which
corresponds to $Q=3,$ and was the original suggestion for a knot
soliton \cite{FN}.

Table 1 shows that there are also trefoil knot solutions for
$Q=8,9,10,11$ and in each case they are obtained from rational map
initial conditions of the type ${\cal K}_{3,2}.$ For $Q=9$ the
trefoil solution is also obtained from a variety of other initial
conditions, including links and the torus knot ${\cal K}_{5,2};$
this last process ${\cal K}_{5,2}\rightarrow{\cal K}_{3,2}$ provides
an example of knot transmutation, where Solomon's seal knot deforms
into a trefoil knot. In Figure \ref{fig-high} a comparison of the 
plots $8{\cal K}_{3,2},9{\cal K}_{3,2},10{\cal K}_{3,2},11{\cal K}_{3,2}$,
emphasizes that the conformation of the knot is important, with each
of these trefoils having a very different structure to the others.
In particular, as the number of twists increases the knot
increasingly contorts and bends, in a manner similar to that seen in
the axial solutions ${\cal A}_{n,1},$ where for $n=3$ the response
to the twist is to lower the energy by bending to break the axial
symmetry. As mentioned earlier, for $n=4,5$ there are similar bent
solutions, which are local minima, where the deformation increases
with $n.$ Trefoil knots therefore appear to follow a similar
pattern, with a minimal number of twists, that is 4, required for
existence and increasingly deformed solutions existing with slightly
larger twists than the minimal value.

The triangles in Figure \ref{fig-energy}, and the associated energy values in
Table 2, show that the ratio of the energy to the bound steadily increases for
the trefoil solutions as the charge increases. It is therefore not
surprising that for a large enough charge, which happens to be
$Q=12,$ a trefoil solution fails to exist. Table 1 confirms that a
$Q=12$ initial condition of the type ${\cal K}_{3,2}$ produces  a
solution which is not a trefoil knot, and will be discussed later.
The triangles in Figure \ref{fig-energy} also confirm that only for the minimal
value $Q=7$ is the trefoil knot the global minimum energy solution,
and in all other cases a trefoil is only a local energy minimum.
However, for $Q=8$ the trefoil knot energy is extremely close to the
lowest energy found, which is a linked solution ${\cal
L}_{3,3}^{1,1}.$ In fact the energies differ by less than $0.1\%,$
which is probably smaller than the accuracy of the computations, so
in this case it is difficult to make a definitive statement
regarding which (if any) is the global minimum.

In Ref.\cite{BS5} a linked solution of the type ${\cal
L}_{2,2}^{2,2}$ was reported, but this appears to be an artefact of
that study being at the limit of computational feasibility for
computing resources available at that time. A configuration of this
type does appear during a relaxation procedure, but after further
relaxation it changes its type. Moreover, an initial condition
of the type ${\cal L}_{2,2}^{2,2}$ can be constructed using the
rational map \be
W=\frac{Z_1^4}{Z_1^4-Z_0^2}=\frac{Z_1^2}{2(Z_1^2-Z_0)}
+\frac{Z_1^2}{2(Z_1^2+Z_0)},\label{map8}\ee and, as seen from Table
1, the relaxation yields a solution of a different type, namely the
linked solution ${\cal L}_{3,3}^{1,1}.$ It therefore now seems
unlikely that a charge eight stable solution exists of the type
${\cal L}_{2,2}^{2,2}.$ Note that the rational map $(\ref{map8})$
describes a link in which both components are linked {\em twice}
with each other, and this is because the irreducible factors of the
denominator are terms like $(Z_1^2-Z_0),$ rather than terms like
$(Z_1-Z_0),$ where the latter corresponds to only single links
between any two components. The partial fraction decomposition in
$(\ref{map8})$ reveals that each component of the link has charge
two, by once again reading off the power of the numerator.

For a given fixed charge there is nothing to prevent multiple
trefoil knot solutions which differ in their conformations and
energies, but no evidence for this phenomenon has been found. For
example, Table 1 shows that the $Q=9$ trefoil knot is produced from
(at least) four different types of initial condition, but in each
case the resulting trefoil solution is the same one with an
identical conformation. This is the case for all the solutions
described in this paper, that is, when a particular solution is
obtained from various initial conditions it is always in the same
conformation, so its charge and type are sufficient to distinguish
it from any other solution.

As the trefoil knot solutions for $Q=9,10,11$ are clearly only local
energy minima then it remains to describe the candidates found for
the global minima for these charges. In fact for all three of these
charges, and also for $Q=12,$ the lowest energy solutions found have
a similar form, which is a link with three components. The simplest
case, $Q=9,$ is displayed in Figure \ref{fig-high} as plot
 $9{\cal L}_{1,1,1}^{2,2,2},$ and as this label suggests there are three
components to the link, each of which has charge one and links once
with each of the other two components. An initial condition 
of the type ${\cal L}_{1,1,1}^{2,2,2}$ with $Q=9$ 
is obtained from the rational map
\be W=\frac{Z_1^3}{Z_1^3-Z_0^3},\label{map9}\ee and quickly relaxes
to the minimal energy solution with the same type. The initial
condition has a more planar arrangement of the three components than
the final solution, and the initial cyclic $C_3$ symmetry is also
broken, as can be seen from plot $9{\cal L}_{1,1,1}^{2,2,2},$
in Figure \ref{fig-high}. Note from Table 1
that this 3-component link is also obtained from a knotted initial
condition of the type ${\cal K}_{4,3},$ but that out of the six
$Q=9$ initial conditions used, four lead to the higher energy trefoil
knot solution. This is evidence that supports the fact that minimal
energy solutions may not be the easiest to find; hence the need to
employ the variety of starting configurations used in this study.

The minimal energy solutions with $Q=10,11,12$ are similar to the
minimal $Q=9$ solution and correspond to increasing one, two and
finally all three, of the charge one components to charge two. In
other words they have the forms ${\cal L}_{1,1,2}^{2,2,2},{\cal
L}_{1,2,2}^{2,2,2}$ and ${\cal L}_{2,2,2}^{2,2,2}$ respectively.
These solutions are displayed in Figure \ref{fig-high}
 and in each case there is
an associated rational map of the same type. Table 1 shows that they
can all be obtained from the relaxation of certain torus knot
initial conditions.

So far the only knot solutions discussed have been trefoil knots. As
the torus knot with the lowest crossing number after the trefoil is
the $(5,2)$-torus knot, then this is the most likely candidate to
appear next as a soliton solution. Table 1 reveals that an initial
condition of the type ${\cal K}_{5,2}$ does not yield a solution of
this type for $Q=8,9,10,$ but it does for $Q=11.$ The $Q=11$
solution is displayed in Figure \ref{fig-high}
 as plot $11{\cal K}_{5,2}.$ It is
only a local energy minimum, despite the fact that it is lower in
energy than the $Q=11$ trefoil knot. There are also solutions of the
type ${\cal K}_{5,2}$ for $Q=12,13,14$ and the pattern mirrors that
of the trefoil knots, in that the ratio of the energy to the 
bound steadily increases with the charge; see the diamonds in 
Figure \ref{fig-energy}.

At $Q=12$ a solution of the type ${\cal K}_{4,3}$ appears (see
Figure \ref{fig-high}
 plot $12{\cal K}_{4,3}$) and its energy is just above that
of the minimal energy ${\cal L}_{2,2,2}^{2,2,2}$ link, though
substantially lower than that of the ${\cal K}_{5,2}$ solution.
Solutions of the type ${\cal K}_{4,3}$ also exist for $Q=13$ and
$Q=14,$ and for these two charges they are the lowest energy
solutions found. The excess energy above the bound is also
reasonably low, so these two solutions are good candidates for the
global minima at $Q=13$ and $Q=14.$  The squares in 
Figure \ref{fig-energy} denote
the ratio of the energy to the bound for ${\cal K}_{4,3}$ solutions, from
which it can be seen that (unlike the other torus knots) 
the lowest charge at which this knot
appears is not the one that is closest to the conjectured bound.

At $Q=13$ a solution exists (see Figure \ref{fig-high} plot $13{\cal X}_{13}$)
which is denoted by ${\cal X}_{13}$ because there is no unambiguous
interpretation as a particular link or knot. This is due to the fact
that a definition of a link or knot requires the position curve not
to self-intersect, but there is nothing to prevent this in the field
theory. For the ${\cal X}_{13}$ solution then either the position
curve self-intersects exactly, or there are parts of the curve
that are so close together that they can not be resolved with the
numerical accuracy currently employed. Some of the lower charge
solutions presented earlier may also appear to self-intersect, for
example the $Q=5$ solution displayed in Figure \ref{fig-low}, but in all these
lower charge solutions the apparent self-intersection can be
resolved by reducing the thickness of the tube plotted around the
position curve, together with a careful consideration of continuity.

If there are no self-intersection points in the solution ${\cal
X}_{13}$ then there are two possibilities for the configuration
type, depending on how this self-intersection point is resolved. The
first possibility is the link ${\cal L}_{1,2,2}^{2,3,3},$ and the
second is a link with two components, where one of the components is
a trefoil knot with five twists and the second component is approximately
axial with a single twist. The latter resolution is essentially 
an intertwining of the charge eight trefoil knot solution with the charge
one solution.
    
 The lowest energy solitons found for $Q=15$
and $Q=16$ are denoted by ${\cal X}_{15}$ and ${\cal X}_{16}$
respectively, and also have points which can not be distinguished from
self-intersection points, as seen from plots $15{\cal X}_{15}$
and $16{\cal X}_{16}$ in Figure \ref{fig-high}.
In each case there is a
resolution into a link with two components where one of the
components is a trefoil knot. The conformation of both the knot
and unknot components in these cases strongly suggests that this 
is the correct resolution, if indeed one is required.
The $Q=16$ solution is the easiest to identify and consists of the 
charge eight trefoil knot intertwined with the minimal energy 
charge two solution, where the conformation of both components is very 
similar to that of the components in isolation.
A field with the correct qualitative behaviour to describe this resolution
is given by the rational map
\be
W=\frac{Z_1^5+Z_1^2Z_0^2+Z_1Z_0^3}{Z_0Z_1^3+Z_0^3}
=\frac{Z_1Z_0^2}{Z_1^3+Z_0^2}+\frac{Z_1^2}{Z_0}.\label{map16}
\ee
Initial conditions generated using this rational map quickly 
leads to the solution ${\cal X}_{16}$ under energy relaxation. 

The solution  ${\cal X}_{15}$ is similar to the solution ${\cal X}_{16},$
except that the resolution involves the charge seven trefoil knot, rather
than the charge eight trefoil knot. 
 The energies of the solutions of type
${\cal X}_n$ are represented by black circles in 
Figure \ref{fig-energy}. 

Note
that $Q=16$ is the lowest charge in which a link with four
components each linking all the others might exist. However, as seen
in Table 1, an initial condition of the type ${\cal
L}_{1,1,1,1}^{3,3,3,3}$ relaxes to the ${\cal X}_{16}$ solution.

The final torus knot solitons presented in this paper are of the
type ${\cal K}_{5,3}$ and exist for $Q=14$ and $Q=15,$ but in both
cases these solutions are only local energy minima. The energies of
these solutions are represented by stars in Figure \ref{fig-energy}.

In summary, it has been shown that there are a variety of torus knot
solitons at various Hopf charges. For example, at $Q=14$ three
different torus knot solutions have been obtained. Most of the knot solitons
found are only local energy minima, but in some cases they are good
candidates for the global minimum.

\begin{figure}
\begin{center}
\leavevmode 
\includegraphics[width=12cm]{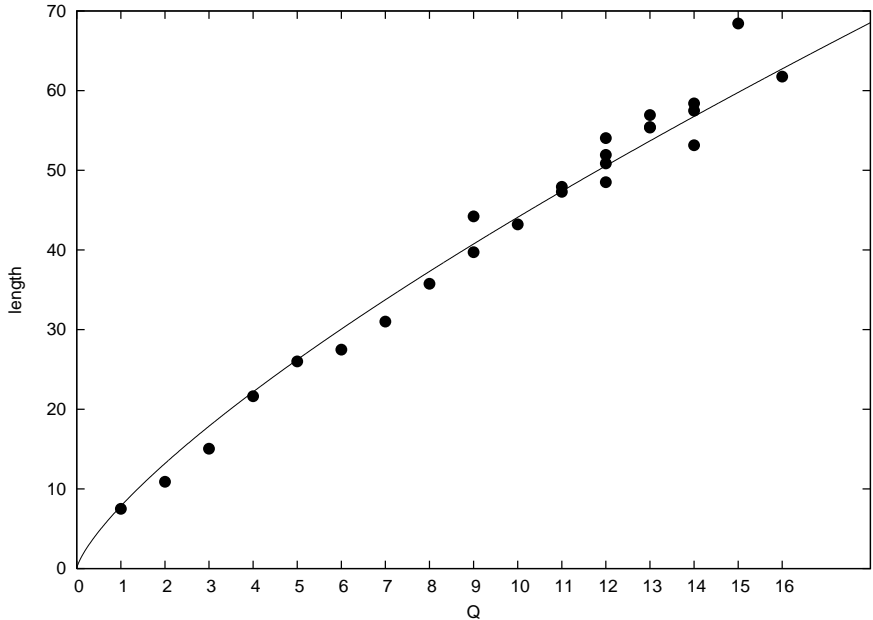} 
\caption{String length (circles) as a function of charge $Q$ for several
solitons, together with the expected growth
behaviour $\gamma Q^{3/4}$ (curve).}
\label{fig-length}
\end{center}
\end{figure}

For a string-like soliton it is expected that one of the 
contributions to the energy should have an interpretation in terms
of the string length. As the soliton energy grows like $Q^{3/4}$
then this suggests that the string length should have a similar growth.
 In Figure \ref{fig-length} the string length is plotted (circles)
for several solitons, including examples of both knots and links.
 Also shown is a curve with the 
expected power growth, $\gamma Q^{3/4},$ where $\gamma$ has been
computed by a least squares fit to be $\gamma=7.86.$ 
This plot demonstrates a reasonable agreement with the expected
$Q^{3/4}$ dependence of the string length. The greatest discrepancy
from the expected string length occurs for the $Q=15$ link 
${\cal L}_{1,1,1}^{4,4,4},$ which is much longer than expected. 
However, note that this solution has an abnormally large ratio of linking
to twist number, with the contribution to the Hopf charge from linking being
four times that from twisting. The solution with the next largest ratio of
linking to twist is the $Q=9$ link ${\cal L}_{1,1,1}^{2,2,2},$ where linking
contributes twice as much as twisting, and the length of this solution
is also a bit above the expected value. It therefore appears that
solutions which link much more than they twist require an increased string
length to minimize energy.   

Recall from the earlier discussion of knot solitons,
that there appears to be a critical twist,
below which knot solitons of a particular type do not exist.
A twist below the critical value seems to be an energetically
inefficient distribution of the Hopf charge between twisting and crossing.
For twists just above the critical value a knot solution continues
to exist but as the twist increases further then again the 
distribution between twisting and crossing becomes inefficient, but this
time due to too much twisting. 

A naive approximation is to assume that for all solutions there is a 
universal optimum twist per unit length. Taking this optimum value from
that of the $Q=1$ soliton, and using the fact that the string length
grows like $Q^{3/4},$ this suggests that an optimum value for the number
of twists is approximately $T\approx Q^{3/4}.$ For a knot with crossing number
 $C$ this gives $C=Q-T\approx Q-Q^{3/4}.$ Using this assumption, a knot
with crossing number $C$ is predicted to be energetically efficient
at an integer charge near to the real value $Q_*,$ which is defined 
as the solution of the equation $Q_*-Q_*^{3/4}=C.$ 
For example, for the trefoil knot
$C=3$ and therefore $Q_*=7.56,$ which roughly agrees with the $Q=7$ and
$Q=8$ trefoils being closest to the energy bound. For $C=5$ then $Q_*=11.07$
and this explains why the $Q=11$ knot ${\cal K}_{5,2}$ is closest to
the bound. For $C=8$ then $Q_*=16,$ so the naive approximation
overestimates the most efficient value, which is seen to be $Q=13$
for the ${\cal K}_{4,3}$ knot.
There are clearly more subtle effects at work than the simple naive
constant twist rate assumed in this calculation, but it does seem to
produce numbers which are in the right ballpark, suggesting that it captures
some qualitative aspects of knot energetics.   

\section{Conclusion}\news
\ \quad The results presented in this paper reveal that there are many
low-energy knot solitons of various types, for a range of Hopf charges, 
together with increasingly complicated linked solitons. The qualitative 
features of these solutions can be replicated using an ansatz involving
rational maps from the three-sphere to the complex projective line, and
futhermore this provides a good supply of initial conditions for
numerical relaxation simulations. The complicated nature of the problem   
means that it is difficult to be certain that the global minimal energy 
solitons have been found, but certainly some good candidates have been 
presented, whose energies agree well with the expected values based on an 
earlier conjectured bound.

The rational map ansatz allows the construction
of any torus knot initial condition, but it is not clear how to extend
this to non-torus knots. All the knot solutions found so far are torus
knots, and it is unknown whether the lack of non-torus knot solitons
is a result of not having suitable initial conditions, or whether there
is some energetic reason to favour torus knots. For example, the figure eight
knot has four crossings and hence a soliton with this form might be expected
with a charge around $Q=9.$ As seen from Table 1, quite a few initial 
conditions have been employed for $Q=9$ and $Q=10,$ and no such solution
has been found. It remains an open problem to understand the absence
(or otherwise) of non-torus knots.

Finally, given that there are many knot and link solitons then it would be
useful if there was an approximate string model that could, even 
qualitatively, reproduce the field theory results. It has been shown that
the string length has the expected behaviour with Hopf charge, so it is
plausible that there may be an approximate description based on a 
string energy that includes contributions from properties of the string such
as its length, twist and writhe, together with relevant interaction terms.

\section*{Acknowledgements}
Many thanks to Michael Farber for helpful discussions.
This work was supported by the PPARC special programme
grant ``Classical Lattice Field Theory''.

\end{document}